\documentclass[a4paper,11pt]{article}
\usepackage{pos}
\usepackage{lineno}
\usepackage{comment}
\title{Recent results on open heavy flavor production ($pp$, $p$Pb, PbPb) from LHCb}

\author*[a]{Chenxi Gu}

\onbehalf{for the LHCb collaboration}

\affiliation[a]{Laboratoire Leprince-Ringuet, École Polytechnique,\\
  91128, Palaiseau, France}

\emailAdd{chenxi.gu@cern.ch}

\abstract{Heavy quarks are produced in the early stages of heavy ion collisions due to their large mass, and subsequently traverse the entire QCD medium evolution. Open heavy flavors provide profound insights into the transport properties of the medium and the process by which quarks neutralize their color charge to form hadrons. In the LHCb experiment, fixed-target collisions cover an unexplored energy range that lies above that of previous fixed-target experiments but below the top RHIC energy for AA collisions. In $p$Pb collisions, heavy quarks are crucial for studying cold nuclear matter effects, which include the modification of nuclear parton distribution functions, energy loss in the nucleus, and other phenomena. These studies provide a baseline for interpreting PbPb measurements.}

\FullConference{%
  31st International Workshop on Deep Inelastic Scattering\\
  8–12 Apr 2024\\
  Maison MINATEC, Grenoble, FRANCE
}

\begin{document}
\maketitle

\section{Introduction}
At hadron colliders, heavy quarks are primarily produced by hard parton-parton interactions in the initial stages of the collisions, and their production is well described by perturbative QCD calculations. These calculations are based on the factorization theorem, which states that heavy-flavor hadron cross-sections depend on the parton distribution functions of the incoming nucleons, the hard parton-parton scattering cross-section, and the fragmentation functions. 

The LHCb detector is a single-arm forward spectrometer that covers a pseudorapidity range of 2 to 5. This unique coverage allows LHCb to probe parton distribution functions at low Bjorken-$x$ ($10^{-5}$). On the other hand, the System for Measuring the Overlap with Gas (SMOG) allows noble gas to be injected in the VErtex LOcator (VELO) to collide with proton or lead beams. This program has extended the probed kinematic range to high Bjorken-$x$, corresponding to the anti-shadowing region.

\section{$D^{0}$ production in $p\mathrm{Ne}$ collisions at \mbox{$\sqrt{s_\mathrm{NN}} = 68.5$} GeV}
\label{Dz}
\begin{figure}[htbp]
\centering
\includegraphics[width=7cm,clip]{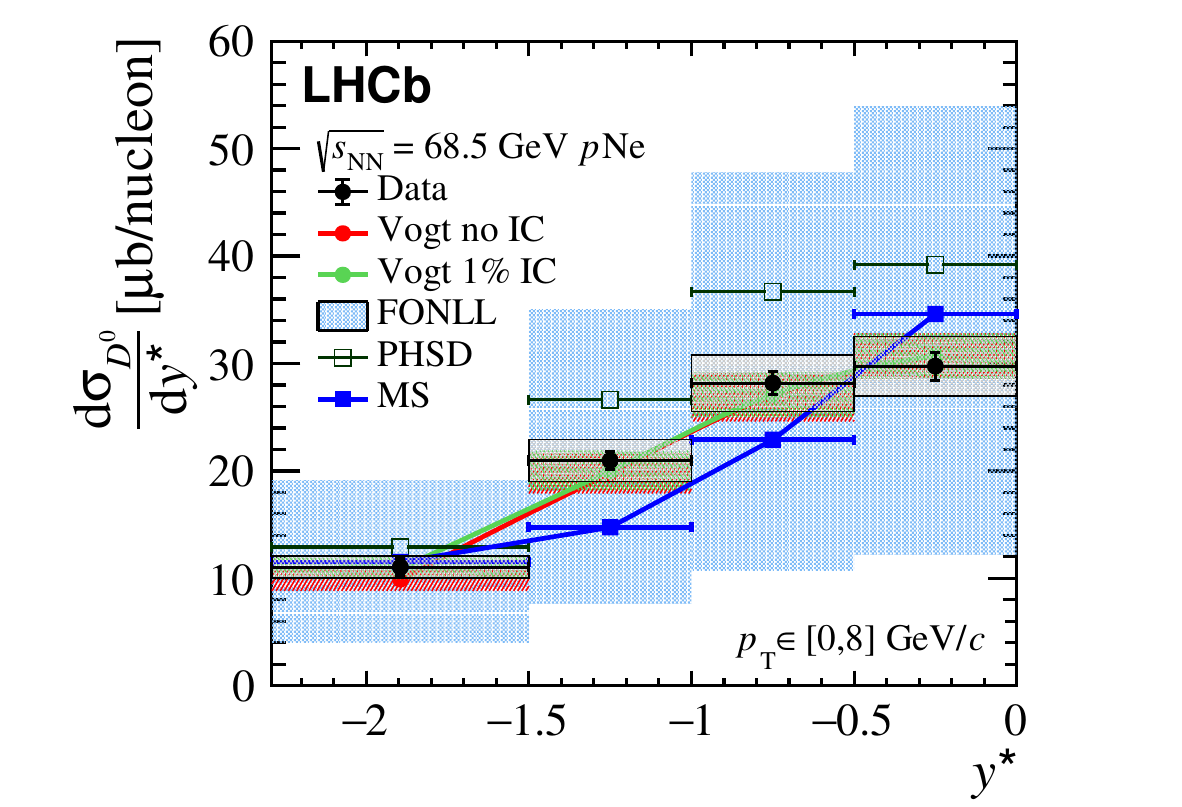}
\includegraphics[width=7cm,clip]{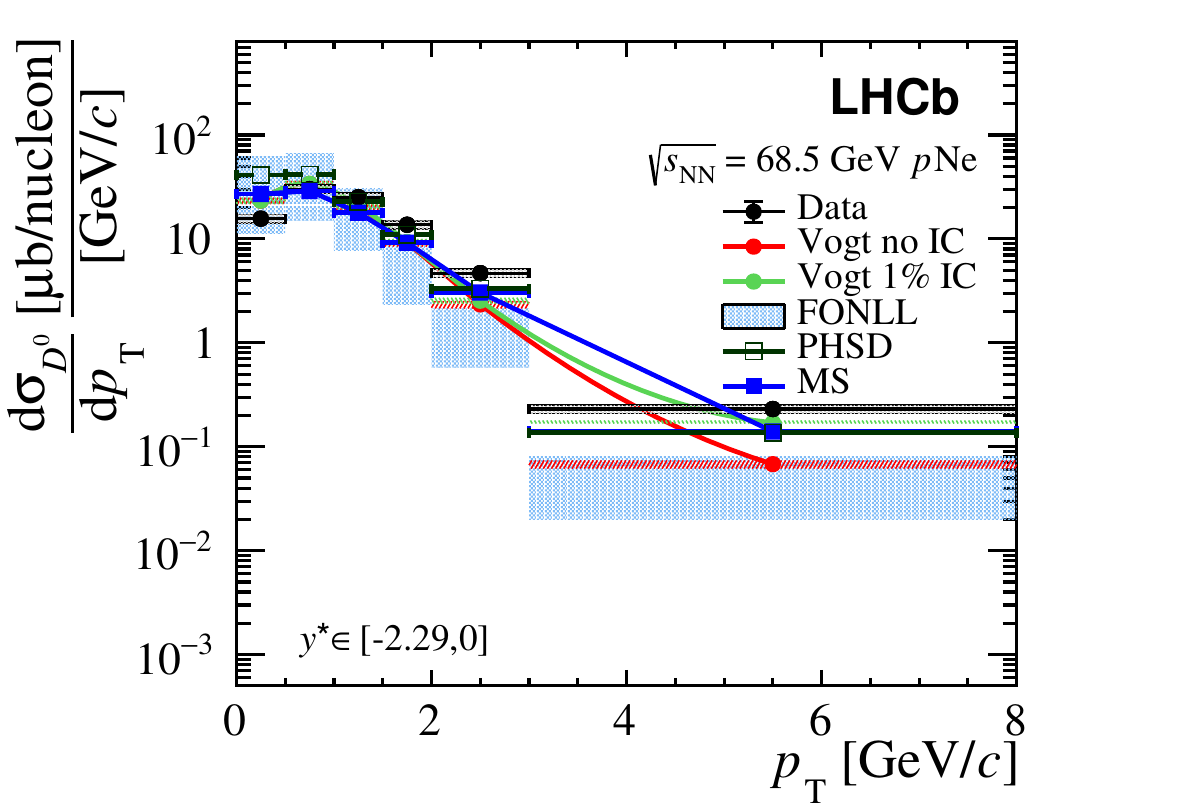}
\caption{Differential cross-section for $D^{0}$ production as a function of $D^{0}$ (left) $y^{*}$ and (right) $p_{\mathrm{T}}$~\cite{LHCb:2022cul}.}
\label{DzpNe}       
\end{figure}

$D^{0}$ meson production in $p$Ne collisions is measured using the 2017 dataset, with an integrated luminosity of $21.7 \pm 1.4$ nb$^{-1}$~\cite{LHCb:2022cul}. Figure~\ref{DzpNe} shows the $D^{0}$ differential cross-sections per target nucleon as functions of $y^{*}$ and $p_{\mathrm{T}}$. These results are compared to theoretical models that incorporate various cold nuclear matter effects. The data points are well matched by alternative predictions, including those with (Vogt 1\% IC) and without (Vogt no IC) intrinsic charm contributions, both of which account for the shadowing effect~\cite{Vogt:2021vsc}. Additionally, predictions (MS) that include 1\% intrinsic charm and 10\% recombination contributions also align closely with the LHCb data. Both FONLL~\cite{Cacciari:1998it,Cacciari:2005rk} and PHSD~\cite{Song:2016rzw} calculations fail to reproduce the $p_{\mathrm{T}}$ distribution, while the rapidity distributions are in better agreement with the data. With more data, distinguishing between models with and without intrinsic charm will become feasible.

\section{$\Lambda_{b}^{0}/B^{0}$ cross-section ratio in high-multiplicity $pp$ collisions at $\sqrt{s} = 13$ TeV}
\label{LbB0}
\begin{figure}[htbp]
\centering
\includegraphics[width=6cm,clip]{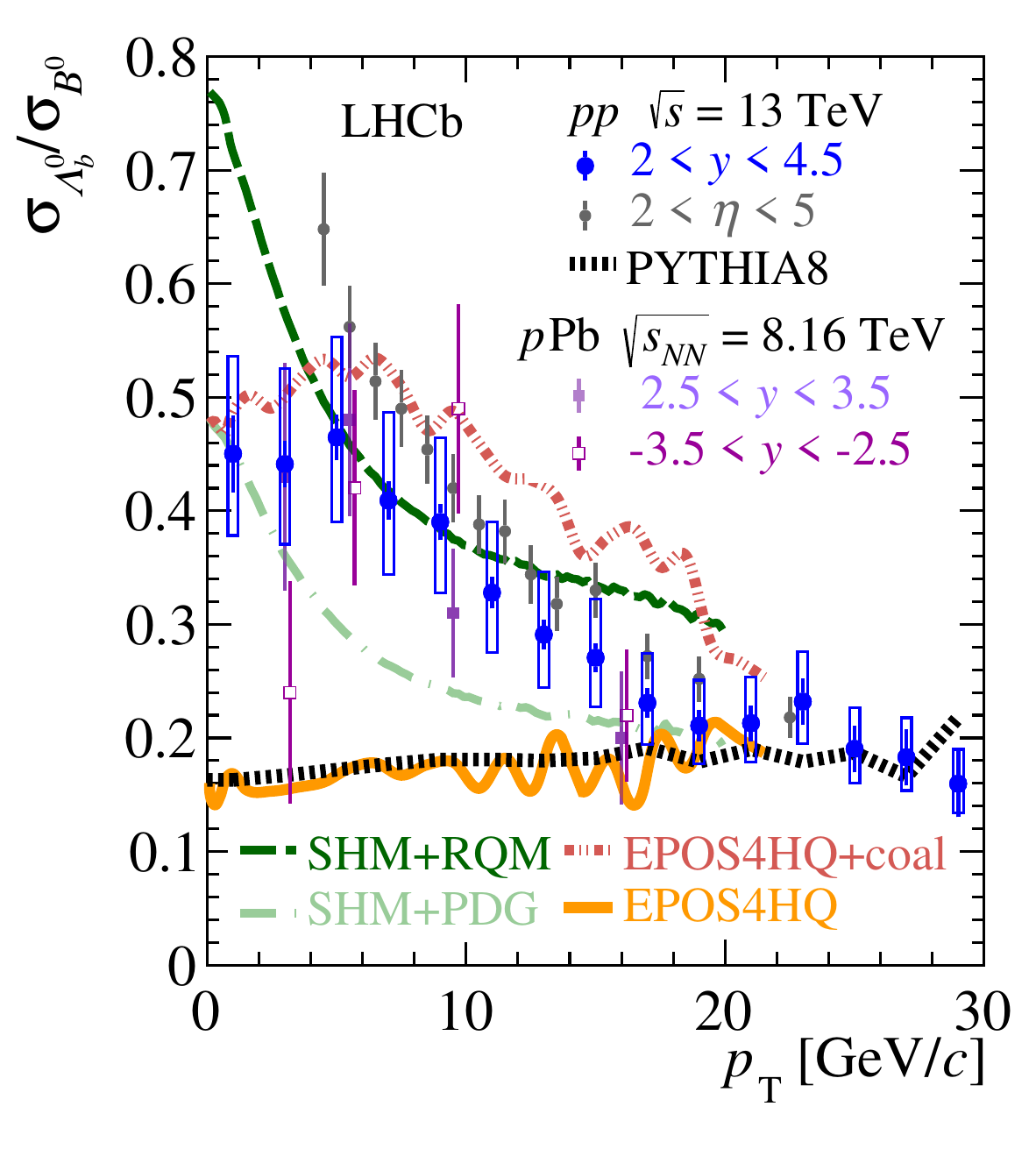}
\includegraphics[width=7cm,clip]{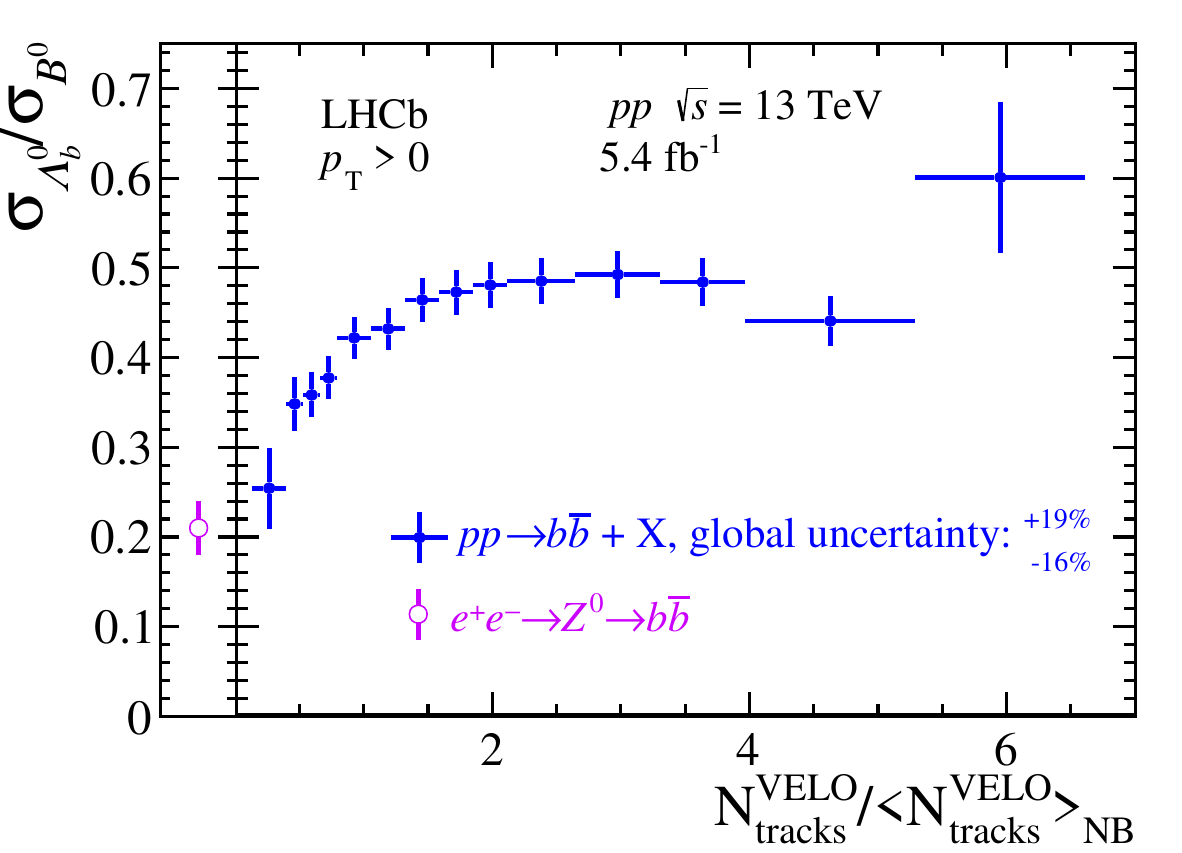}
\caption{Cross-section ratio $\sigma_{\Lambda_{b}^{0}}/\sigma_{B^{0}}$ as a function of (left) $p_{\mathrm{T}}$ and (right) normalized multiplicity~\cite{LHCb:2023wbo}.}
\label{LbB0PT}       
\end{figure}

The left plot of Figure~\ref{LbB0PT} presents the $\Lambda_{b}^{0}/B^{0}$ cross-section ratio as a function of $p_{\mathrm{T}}$ in $pp$ collisions at a center-of-mass energy of $\sqrt{s} = 13$ TeV~\cite{LHCb:2023wbo}. The data are compared to previous $pp$ measurements~\cite{LHCb:2019fns} and $p$Pb measurements~\cite{LHCb:2019avm}, and are generally consistent with them within uncertainties. Additionally, calculations from the $b$ quark statistical hadronization model~\cite{He:2022tod} and EPOS4HQ~\cite{Zhao:2023ucp} are included. The light green dashed curve considers feeddown contributions from $b$ baryons collected by the Particle Data Group~\cite{ParticleDataGroup:2020ssz}. The dark green dashed curve accounts for feeddown contributions from an expanded set of $b$ baryons predicted by the Relativistic Quark Model~\cite{Ebert:2011kk}. By incorporating a coalescence mechanism, the EPOS4HQ model provides a more accurate description of the data.

The right plot of  Figure~\ref{LbB0PT} presents the $\Lambda_{b}^{0}/B^{0}$ cross-section ratio as a function of the normalized multiplicity. The event multiplicity is parametrized by $\mathrm{N}_{\mathrm{tracks}}^{\mathrm{VELO}}$. 
$\mathrm{N}_{\mathrm{tracks}}^{\mathrm{VELO}}$ denotes the total number of charged tracks reconstructed in the VELO detector.
The plot shows that the $\Lambda_{b}^{0}/B^{0}$ ratio increases significantly with multiplicity. In the lowest multiplicity bin, the $\Lambda_{b}^{0}/B^{0}$ ratio reaches a value comparable to that observed in $e^{+}e^{-}$ collisions.

\section{$D^{+}$ and $D_{s}^{+}$ production in $p\mathrm{Pb}$ collisions}
\label{DsDp}
The LHCb experiment measures the prompt production of $D^{+}$ and $D_{s}^{+}$ in $p$Pb collisions at \mbox{$\sqrt{s_\mathrm{NN}} = 5.02$} TeV~\cite{LHCb:2023kqs} and \mbox{$\sqrt{s_\mathrm{NN}} = 8.16$} TeV~\cite{LHCb:2023rpm}. The measurements are performed in two collision configurations: forward collisions, where the proton beam is directed towards the LHCb detector, and backward collisions, where the lead beam is directed towards the LHCb detector.

\begin{figure}[htbp]
\centering
\includegraphics[width=7cm,clip]{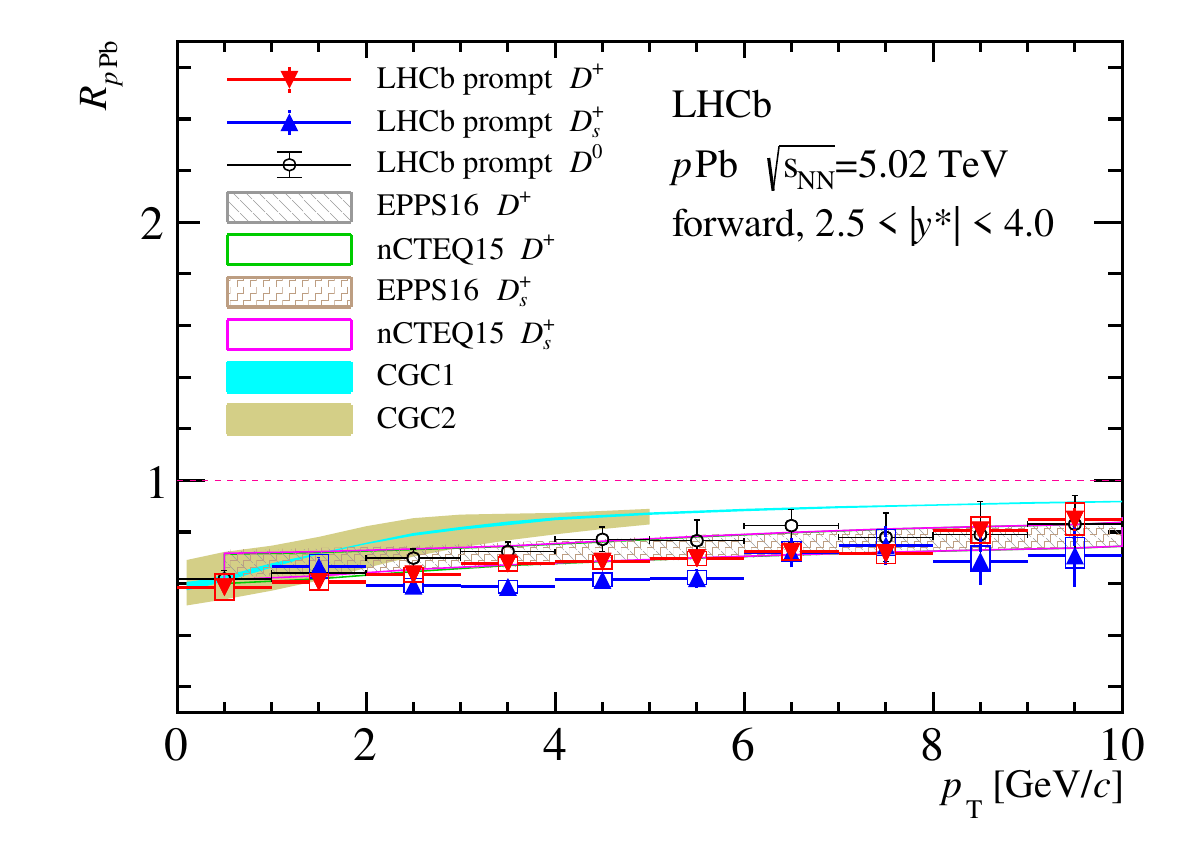}
\includegraphics[width=7cm,clip]{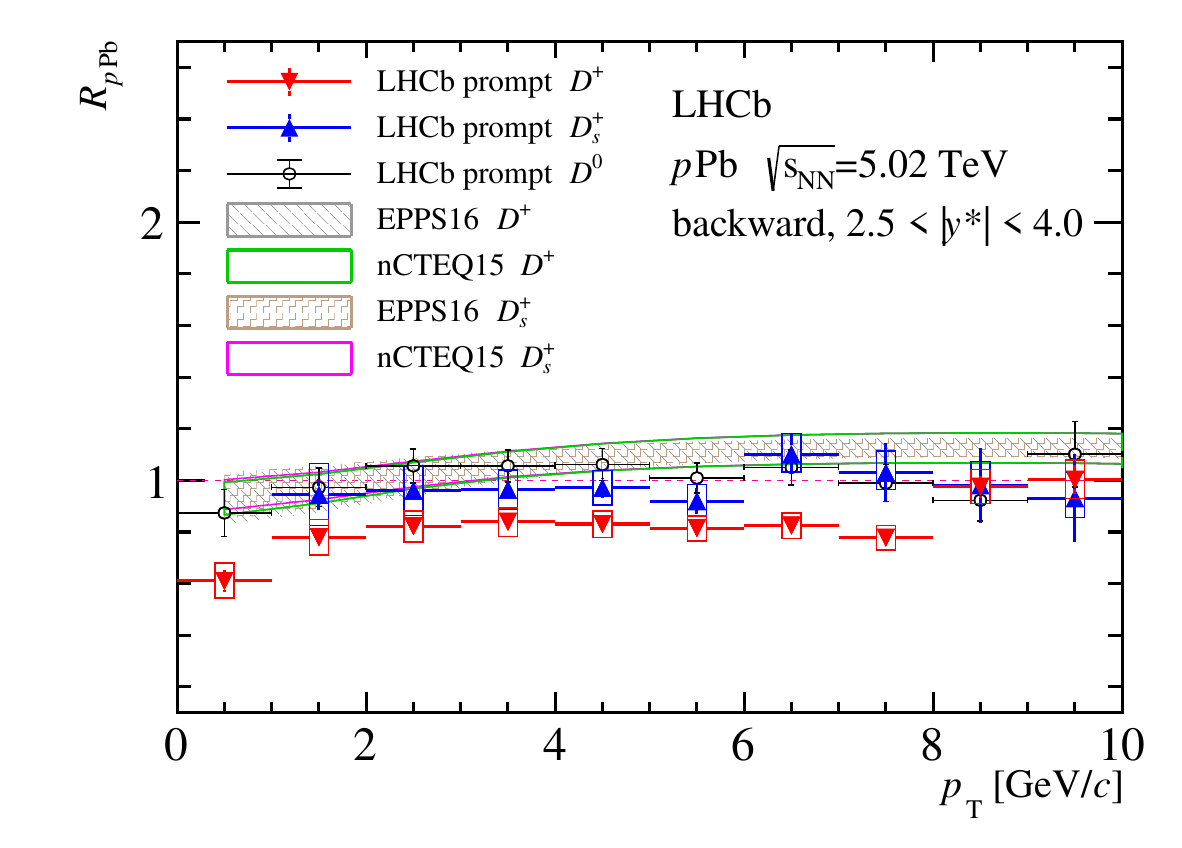}
\caption{Nuclear modification factor of $D$ mesons in $p$Pb collisions at \mbox{$\sqrt{s_\mathrm{NN}} = 5.02$} TeV~\cite{LHCb:2023kqs} at forward rapidity (left) and backward rapidity (right)}
\label{DsDpProduction}       
\end{figure}

\begin{figure}[htbp]
\centering
\includegraphics[width=13cm,clip]{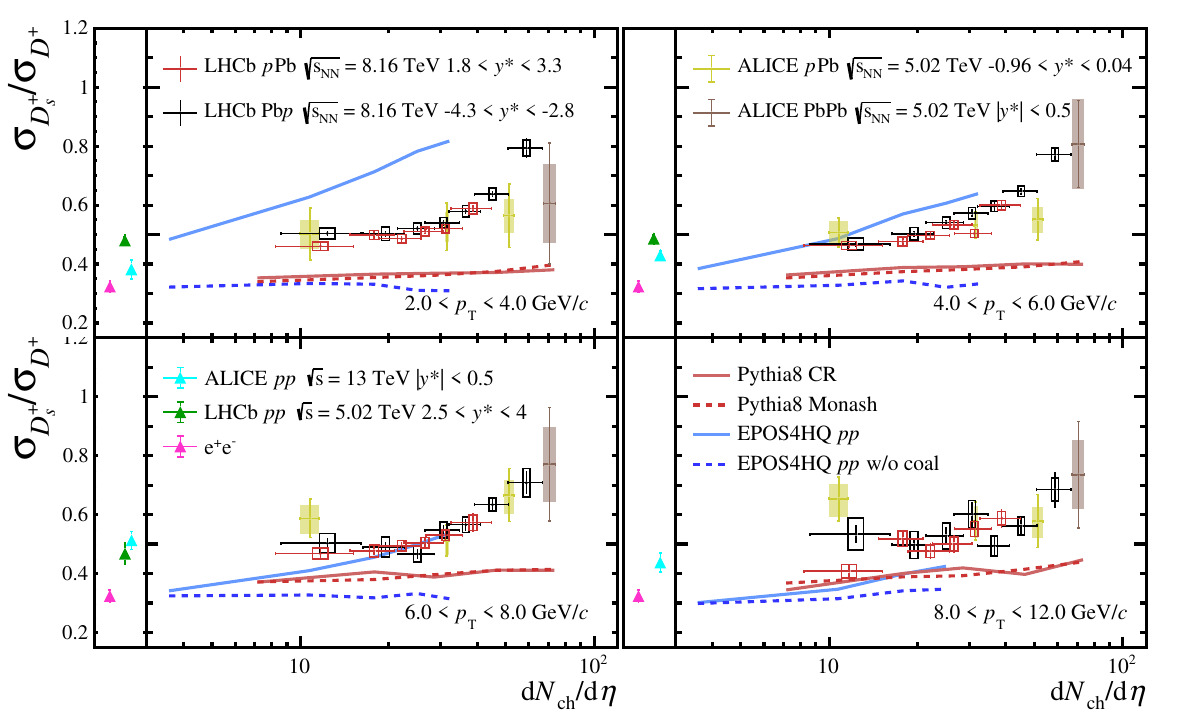}
\caption{The cross-section ratio $D_{s}^{+}/D^{+}$ as a function of $\text{d}N_{\text{ch}}/\text{d}\eta$ in $p\mathrm{Pb}$ collisions at $\sqrt{s_\mathrm{NN}} = 8.16$ TeV is studied for different $p_{\mathrm{T}}$ ranges: 2 < $p_{\mathrm{T}}$ < 4 GeV/$c$ (top left), 4 < $p_{\mathrm{T}}$ < 6 GeV/$c$ (top right), 6 < $p_{\mathrm{T}}$ < 8 GeV/$c$ (bottom left), and 8 < $p_{\mathrm{T}}$ < 12 GeV/$c$ (bottom right)~\cite{LHCb:2023rpm}.}
\label{DsDpRatio}       
\end{figure}

Figure~\ref{DsDpProduction} presents the nuclear modification factor, $R_{p\text{Pb}}$, for $D$ mesons in $p$Pb collisions at \mbox{$\sqrt{s_\mathrm{NN}} = 5.02$} TeV. The $R_{p\text{Pb}}$ is consistent with nPDFs~\cite{Eskola:2016oht,Kovarik:2015cma} and CGC~\cite{Ducloue:2015gfa} calculations in the forward collisions, while in backward collisions, the $R_{p\text{Pb}}$ for $D^{+}$ is observed to be lower than the theoretical predictions.
Figure~\ref{DsDpRatio} illustrates the $D_{s}^{+}/D^{+}$ cross-section ratio as a function of the charged particle density, $\text{d}N_{\text{ch}}/\text{d}\eta$, in $p\mathrm{Pb}$ collisions at $\sqrt{s_\mathrm{NN}} = 8.16$ TeV~\cite{LHCb:2023rpm}. 
The ratio of $D_{s}^{+}/D^{+}$ increases with $\text{d}N_{\text{ch}}/\text{d}\eta$ across all $p_{\mathrm{T}}$ intervals, showing a similar pattern in both forward and backward collisions, which indicates that this ratio is independent of rapidity and strongly correlated with charged particle density.
Theoretical calculations using EPOS4HQ~\cite{Zhao:2023ucp,Zhao:2024ecc} are included for comparison. Although there are some discrepancies with experimental data, EPOS4HQ effectively captures the multiplicity-dependent trends across all $p_{\mathrm{T}}$ intervals.

\section{Summary and outlook}
\label{summary}
These measurements of open heavy-flavor provide new insights into nuclear structure and the mechanism of heavy-quark hadronization. For Run 3, the fixed-target experiment has been upgraded with the SMOG2 gas storage cell, which increases local gas pressure by up to two orders of magnitude and allows for parallel data-taking with collider mode, resulting in an enhanced luminosity. Additionally, thanks to the upgrade of all tracking detectors in LHCb to higher granularity, the centrality of reconstructed PbPb events can be pushed to 30\%. Our research will benefit greatly from these upgrades.

\bibliographystyle{JHEP}
\bibliography{my-bib-database}

\acknowledgments
This work has received funding from the European Union's Horizon 2020 research and innovation programme under the Marie Sk{\l}odowska-Curie grant agreement No.~899987 (EuroTechPostdoc2).

\end{document}